\documentclass[a4paper]{article}
\pdfoutput=1
\usepackage{pdfpages}
\usepackage{icrc2013}
\usepackage{amsmath}
\usepackage[english]{babel}
\usepackage[utf8]{inputenc}
\usepackage[T1]{fontenc}
\usepackage{cite}
\usepackage{url}
\usepackage{hyperref}
\hypersetup{
  colorlinks=true,
  linkcolor=blue,
  citecolor=blue,
  urlcolor=blue,
}
\title{The High Altitude Water Cherenkov Observatory}
\shorttitle{The HAWC Observatory}
\authors{
Miguel A.\ Mostaf\'{a}$^{1,2}$
for the HAWC Collaboration.
}
\afiliations{
$^1$ Department of Physics, Colorado State University, Ft Collins, CO 80523, USA\\
\scriptsize{
$^{2}$ now at The Pennsylvania State University, University Park, PA 16802, USA.}}
\email{miguel@psu.edu}
\abstract{The High Altitude Water Cherenkov (HAWC) observatory is a large field of view, continuously operated, TeV $\gamma$-ray
experiment under construction at 4,100~m a.s.l.\ in Mexico. The HAWC observatory will have an order of magnitude better
sensitivity, angular resolution, and background rejection than its predecessor, the Milagro experiment. The improved
performance will allow us to detect both transient and steady emissions, to study the Galactic diffuse emission at TeV
energies, and to measure or constrain the TeV spectra of GeV $\gamma$-ray sources. In addition, HAWC will be the only
ground-based instrument capable of detecting prompt emission from $\gamma$-ray bursts above $50$~GeV.
The HAWC observatory will consist of an array of 300 water Cherenkov detectors (WCDs), each with four
photomultiplier tubes. This array is currently under construction on the flanks of the Sierra Negra volcano near the city of
Puebla, Mexico. The first thirty WCDs (forming an array approximately the size of Milagro) were deployed in Summer
2012, and 100 WCDs will be taking data by May, 2013. 
We present in this paper the motivation for constructing the
HAWC observatory, the status of the deployment, and the first results from the constantly growing array.}

\begin{document}
\maketitle

\section{Introduction}

The High Altitude Water Cherenkov (HAWC) observatory is a survey instrument capable of 
expanding the catalog of TeV $\gamma$-ray sources, and of 
monitoring most of the objects in this catalog for flaring activity. 
The HAWC observatory will not only detect new sources, 
but will also promptly initiate campaigns of 
multi-messenger observations. 

The main objectives of the HAWC collaboration are to contribute to the study of
the origin of cosmic rays via its observations of $\gamma$-rays up to 100~TeV from discrete sources and the Galactic plane,
to understand particle acceleration in astrophysical jets using HAWC's wide field of view and high duty factor observations of transient sources (such as $\gamma$-ray bursts and supermassive black holes), and
to explore new TeV physics via HAWC's unbiased survey of 
the sky.

The deployment of water Cherenkov detectors started in Summer 2011, 
and $1/3$ of the final array was already completed by the time of this conference. 
We present in this paper the design of the observatory, the current status of the deployment, 
and the first results using data taken with a small fraction of the final layout.

\section{Design and operation of HAWC}
The HAWC design builds upon the experience with the Milagro detector. 
Milagro was the first generation of $\gamma$-ray detectors using the water Cherenkov technology. 
Instead of the double layer of PMTs used in Milagro, 
the HAWC design utilizes a single deep layer of PMTs with a wider separation. 
This configuration gives HAWC a much larger active area than Milagro for the same photo-cathode area.
Another important different with respect to Milagro is that instead of a big pond 
the HAWC observatory consists of a large array of water Cherenkov detectors (WCDs).

\begin{figure}[!htb]
  \centering
  \includegraphics[width=0.425\textwidth]{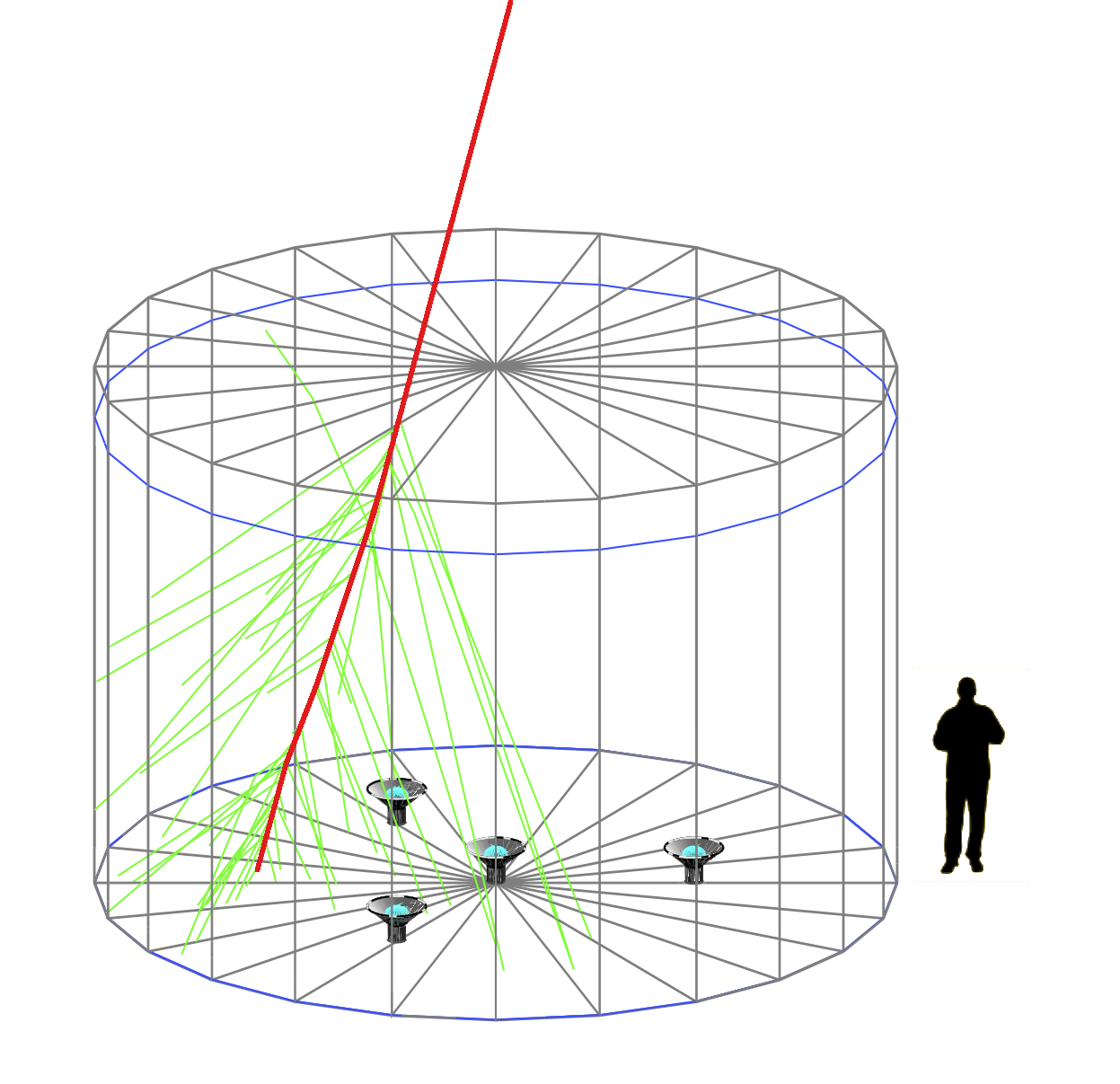}
  \caption{Simulation of a 1~GeV muon (red line) passing through a HAWC detector and emitting Cherenkov light (green lines). The layout of the four PMTs is shown, and the human silhouette provides the scale of the water Cherenkov detectors.}
  \label{fig:wcd}
 \end{figure}

The structural support of the detectors is provided by corrugated steel tanks of $7.3$~m diameter.
There is a light-tight plastic bladder inside each tank.
Bladders are filled with ultra-pure water to a level of $4.5$~m.
There are four upward-facing, baffled photomultiplier tubes (PMTs) in each WCD anchored to the bottom of the tank. 
Three of the four PMTs 
are $8"$ hemispherical \texttt{Hamamatsu R5912} (previously used in the Milagro experiment) 
spaced $6$~ft from the center of the tank.
The fourth PMT, positioned in the center, is a high quantum-efficiency $10"$ PMT designed to increase the efficiency of the observatory to low-energy showers.
We show a simulation of a single air-shower muon penetrating one of the WCDs in Fig.~\ref{fig:wcd}. 

The HAWC observatory will consist of an array of 300 WCDs to observe particles from air showers. 
Hence, the observatory will comprise 1200 PMTs in total. 
The optical isolation of the PMTs between detectors will reduce the noise rate. 
This is an important consideration at high altitude due to the larger flux of low energy cosmic rays.

\begin{figure}[!ht]
  \centering
  \includegraphics[width=0.45\textwidth]{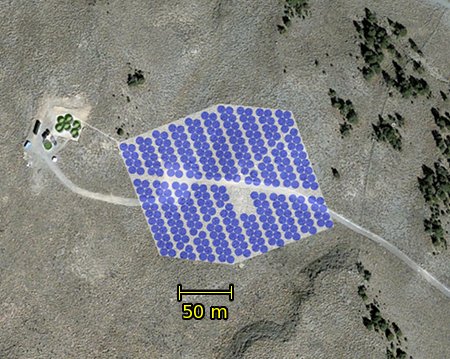}
  \caption{Relative positions of the 300 water Cherenkov detectors (blue circles) that will comprise the HAWC array. The positions of the tanks in the VAMOS prototype are also shown (green circles).}
  \label{fig:layout}
 \end{figure}

Before the deployment of the production detectors, we tested the design and concept at the site with a small engineering prototype.
The prototype array was called VAMOS and consisted of six WCDs.
We show in Fig.~\ref{fig:layout} the layout of the HAWC array and the 6-tank VAMOS prototype. 
(The seventh tank in the prototype was used for water storage.) 
The empty area in the center of the array is the location of the central electronics building. 
Readout cables will run from the central building to the WCDs via the channel which runs from west to east across the array.
WCDs are being deployed in a dense pattern that, when completed, 
will provide $62\%$ coverage of the $135$~m $\times150$~m instrumented area.  

\begin{figure*}[!hbt]
  \centering
  \includegraphics[width=0.8\textwidth]{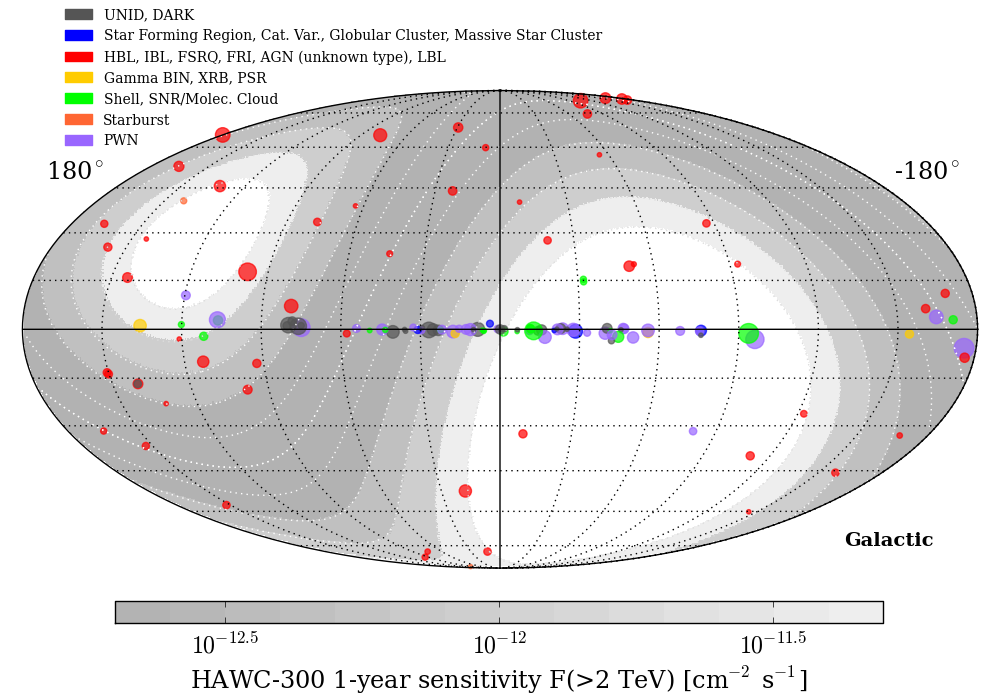}
  \caption{The sky coverage of the HAWC observatory in Galactic coordinates. The detector sensitivity is shown in the gray bands as a function of declination. 
  Sources of very high energy $\gamma$-rays are from the TeVCat online catalogue~\cite{bib:TeVCat}.}
  \label{fig:sensimap}
 \end{figure*}

As a wide field of view instrument, the HAWC observatory is well suited to address the open question of cosmic-ray origins 
through the measurement of diffuse and extended TeV $\gamma$-ray emission from large-scale areas in our Galaxy. 
HAWC will provide an unbiased survey of a large portion of the Northern Hemisphere down to regions close to the Galactic center 
at energies above $100$~GeV.
We show in Fig.~\ref{fig:sensimap} the sky visible to the HAWC observatory and the 
dependence with declination of the detector sensitivity. 
The HAWC data will allow for the study of diffuse emission from large areas along the Galactic plane 
and from structures such as the Fermi bubbles.

\subsection{Laser calibration}
The relative times between the signals in the PMT channels 
have to be calibrated with nanosecond precision over the full dynamic range of the PMTs 
to properly reconstruct the arrival directions of the primary particles. 
The response time of a PMT and electronics depends on the light intensity striking on the PMT. 
A laser calibration system was designed to accurately measure the relative timing among the PMT channels. 
Short ($300$~ps) laser pulses with varying intensities are sent into every WCD through optical splitters, switches, and optical fibers. 
The time between the laser shot and the PMT signal is recorded to correct for the dependence on the light intensity. 
A time residual study is also performed to improve the angular reconstruction. 
The time residual is the difference between a fitted air shower front and the PMT readout time. 
This systematic time offset is then accounted for in an iterative shower reconstruction procedure to improve the determination of the incoming direction of the primary particle. 
In Ref.~\cite{bib:Hugo} we present the first results of the timing calibration curves and time residual studies.

\begin{figure*}[!ht]
  \centering
  \includegraphics[width=0.5\textwidth]{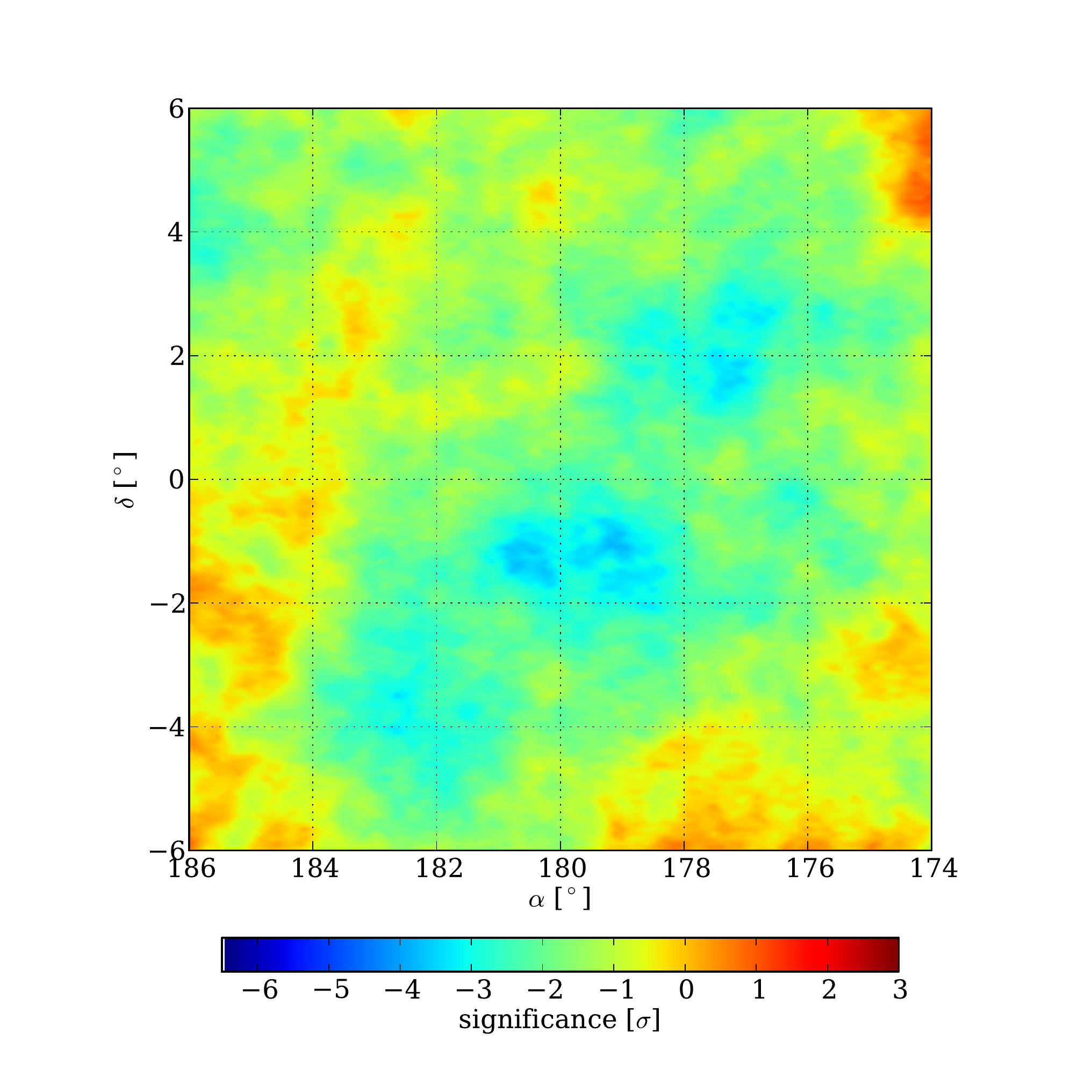}%
  \includegraphics[width=0.5\textwidth]{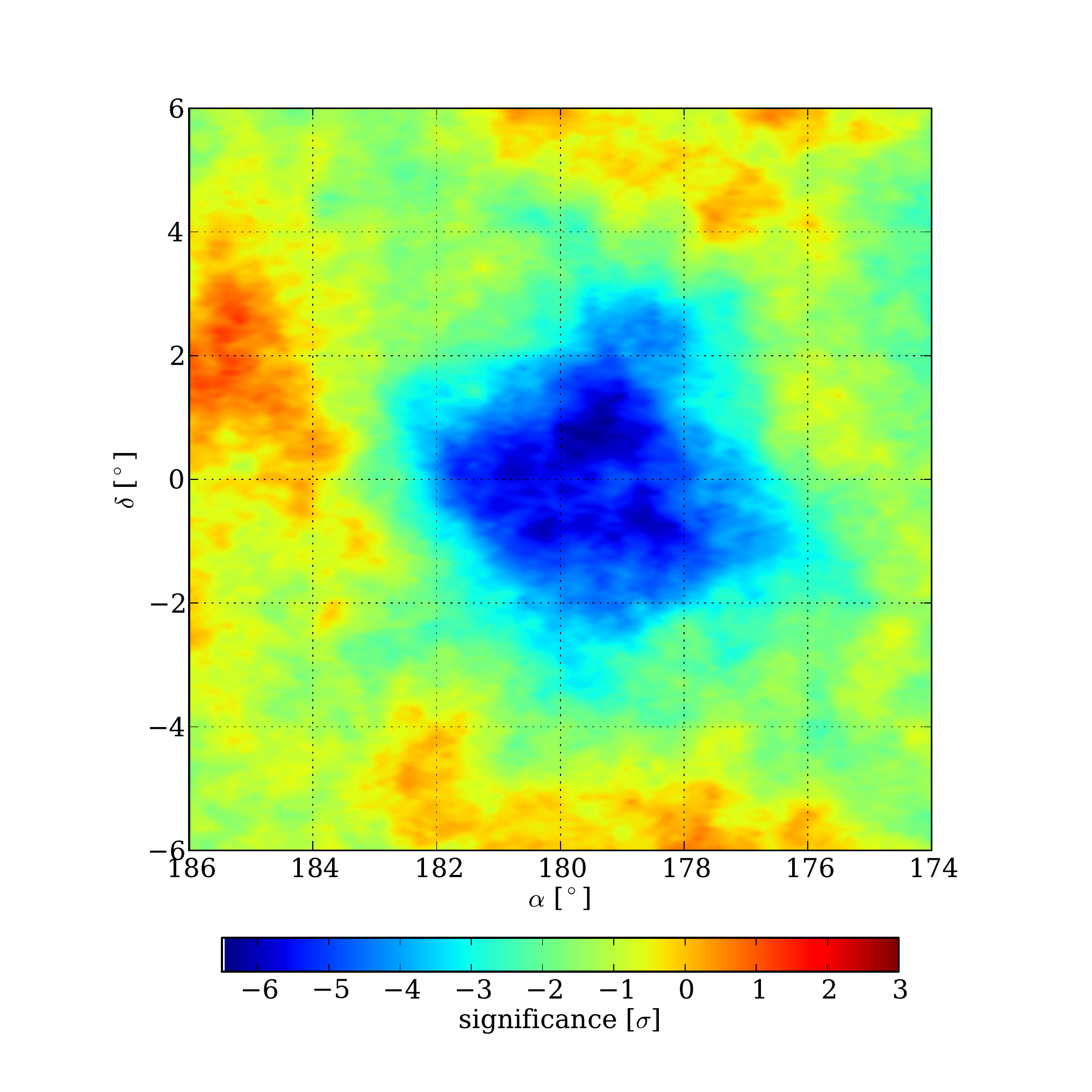}
  \caption{Significance maps of air shower event directions centered on the position of the moon, where $\alpha =$ event right
ascension $(RA) -$ moon $RA + 180^{\circ}$ and $\delta =$ event declination $-$ moon declination. The map on the left side is based on reconstructions
with all pulse charges set to $1$~photoelectron and no timing calibration. The map on the right side was produced with full charge and
timing calibrations included.
The time and charge calibrations are explained in Refs.~\cite{bib:Hugo,bib:Robert}.}
  \label{fig:calib}
 \end{figure*}

As the laser calibration system provides intensities from $0.1$~photo-electrons (PEs) to over $1000$~PEs, 
the charge calibration of all PMTs is also obtained using the laser pulses,
and it further improves the angular resolution of measured arrival directions.
In Ref.~\cite{bib:Robert} we describe the design and performance of the laser calibration system. 
We show in Fig.~\ref{fig:calib} 
the improvement in the geometrical reconstruction of air showers due to the (time and charge) calibration.
These effects are verified using the measurement of the shadow that the moon produces in the flux of cosmic rays.
The results that we show in Fig.~\ref{fig:calib} are
based on data collected with a partial array of 30 WCDs.

The main components of the HAWC laser calibration system are installed and operational at the site in Mexico.
Calibration runs are performed regularly without adding significant dead time. 
The experience gained from a WCD prototype in Colorado, and the
systematic time residuals derived from shower fits made it possible to calibrate the HAWC data even before calibration
results for all individual PMT channels were available. 
Both charge and timing calibrations significantly improve the reconstruction of the arrival directions of air showers. 
The network of optical fibers  will continue to grow together with the HAWC array, 
The dedicated laser system provides the regular calibration of all PMTs to guarantee a stable performance of the array.

\subsection{Site and deployment status}

The HAWC observatory is being built at an altitude of 4,100~m at the Sierra Negra volcano in the state of Puebla, Mexico.
This altitude corresponds to an atmospheric depth of $\sim640$~g$/$cm$^2$.
This closer proximity to the shower maximum allows, in comparison to Milagro, 
twice as many particles in a shower at the same energy, and a lower energy threshold.
The HAWC site is located at 19$^\circ$ North and $97^\circ$ West.
This allows for observations of $2/3$ of the sky (restricting observations to within 45$^\circ$ of the local zenith), 
including a significant overlap with the HESS survey of the Galactic plane, and with the fields of view of IceCube and VERITAS.
HAWC will have considerable daily exposure to the Crab, Geminga, Mrk~$421$, Mrk~$501$, and the Cygnus region.
HAWC will also be able to reach the Galactic Center at $46^\circ$ from the zenith.

Installation work started at the HAWC site in Spring 2010 with the deployment of the 
VAMOS engineering prototype~\cite{bib:VAMOS}.
This prototype array consisted of six WCDs with a total of 36 PMTs.
The location of the prototype array with respect to the HAWC array can be seen in Fig.~\ref{fig:layout}.
We operated the VAMOS prototype array between September, 2011 and June, 2012. 
We present a more detailed description of the prototype and the first results in Ref.~\cite{bib:VAMOS}.
The experience with this prototype was useful to validate working at the high altitude site, but 
also to optimize and finalize the design of the different components of the WCDs.

\begin{figure}[!ht]
  \centering
  \includegraphics[width=0.45\textwidth]{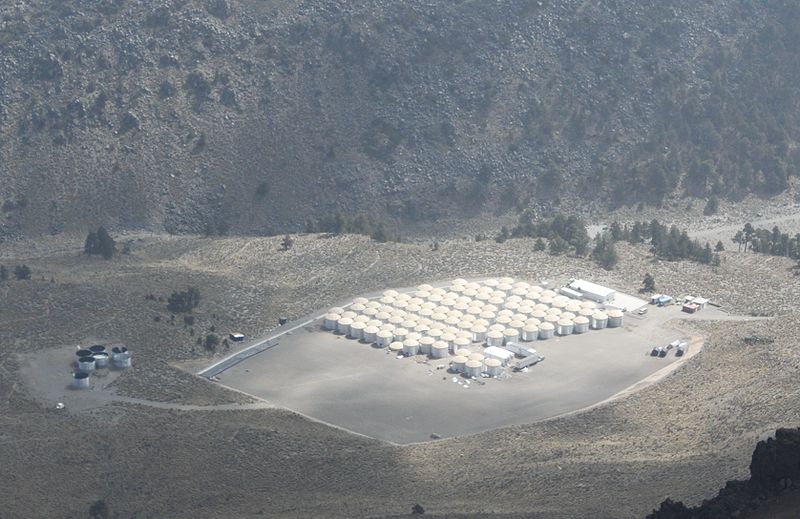}
  \caption{Status of the deployment of the HAWC array on May 16, 2013. The
cleared area indicates the size of the completed instrument, $\sim$22,000~m$^2$.}
  \label{fig:status}
 \end{figure}

The deployment of WCDs at the HAWC site started in March, 2012. 
By August, 2012 we started operating the first 30 WCDs. 
This sub-array, referred to as HAWC-30, has a comparable area to the Milagro experiment.
We used an average calibration for the events recorded with HAWC-30.
The laser calibration system~\cite{bib:Robert} started running at the HAWC site in Summer 2013.
We show in Fig.~\ref{fig:status} the status of the HAWC array by May, 2013.

\begin{figure}[!ht]
  \centering
  \includegraphics[width=0.4\textwidth]{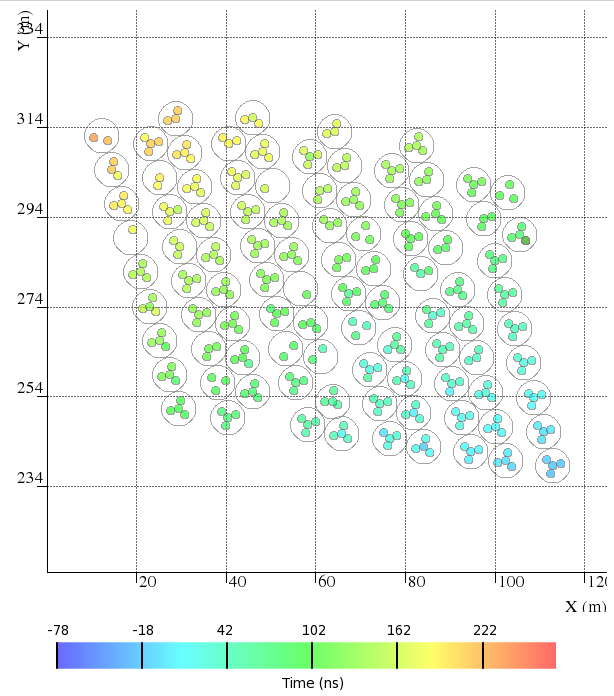}
  \caption{First 95-detector (341-PMT) event recorded with HAWC on June 12, 2013.}
  \label{fig:hawc95}
 \end{figure}

We started taking data with an array of 95 WCDs (HAWC-95) on June 12, 2013. 
We show in Fig.~\ref{fig:hawc95} an event with signal in all operational WCDs on the first day of HAWC-95 operation.
In this conference we present results using data from the VAMOS, HAWC-30, and HAWC-95.

Full operation of HAWC-111 started in September, 2013. 
Currently, the PMTs are being calibrated in situ, and we use the calibration curves corresponding to each individual PMT.
The HAWC-250 array is on schedule for completion in August, 2014, 
and full HAWC observatory will be completed by the end of 2014.

We described in more detail 
the HAWC site, the site infrastructure, layout and platform, and the design and construction of the WCDs
in Ref.~\cite{bib:Ibrahim}.

\section{Sensitivity to $\gamma$-ray sources}

The design of the HAWC observatory combines the Milagro water Cherenkov technology with a high altitude site. 
Re-deploying the existing Milagro photomultiplier tubes (PMTs) and electronics in a different configuration 
at an altitude of 4,100~m will lead to a sensitivity increase of a factor of $\sim15$ over Milagro. 
This improvement is due to the higher altitude, the increased physical area, and the optical isolation of the PMTs. 
As a result, the HAWC detector will see a $5\sigma$ signal from the Crab Nebula in a single $4$-hr transit 
(compared to $\sim5$ months for Milagro)
while maintaining this sensitivity over $2\pi$~sr. 

\begin{figure}[!ht]
  \centering
  \includegraphics[width=0.5\textwidth]{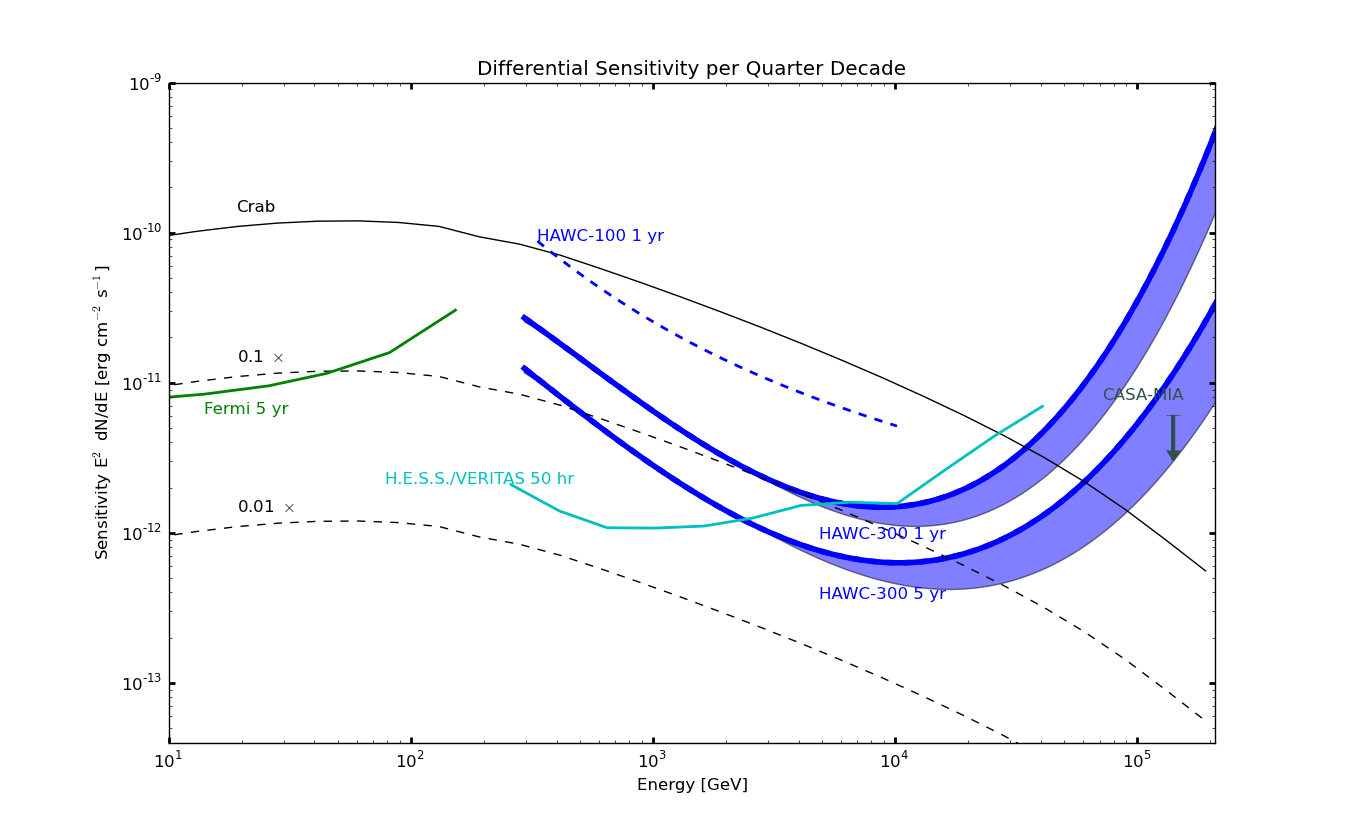}
  \caption{Differential sensitivity per quarter decade of HAWC for 1 and 5 years (and 1 year of HAWC-100) is shown compared to other existing and future IACTs. 
  Note that the sensitivity of HAWC (and Fermi-LAT) is for a sky survey while it is for 50 hours on a source for IACTs.}
  \label{fig:sensitivity}
 \end{figure}

We show in Fig.~\ref{fig:sensitivity} the differential sensitivity of HAWC as a function of energy.
The HAWC observatory will be used to study particle acceleration in Pulsar Wind
Nebulae, Supernova Remnants, Active Galactic Nuclei and $\gamma$-ray Bursts.
We describe in Ref.~\cite{bib:Jordan} how the sensitivity and its uncertainty are calculated.

The Cygnus region is an extremely active star forming region with a
wealth of $\gamma$-ray sources such as pulsar wind nebulae, young star
clusters, and binary systems.
Observations from radio to TeV
energies have revealed diffuse emission and a variety of objects
such as molecular clouds, star clusters, and pulsars. 
The Milagro experiment
detected two extended sources in the TeV regime (MGRO\,J2019+37 and J2031+41) along
with hints of correlated GeV emission in the region. 
The HAWC observatory will have better angular and energy resolution, and one order of magnitude more sensitivity than the
Milagro experiment at similar energies. 
Thus, data from HAWC will allow us to resolve the TeV emission from this region to study the morphology of the sources,
and also to improve the spectral measurements to investigate the origins of the emission in this region.
We present in Ref.~\cite{bib:Michelle} the sensitivity of the HAWC observatory to the Cygnus region.

Measurements of the very high energy diffuse $\gamma$-ray emission are an excellent probe of cosmic-ray acceleration, propagation, and density distribution at different locations within our Galaxy. 
At TeV energies, the Milagro collaboration reported a significant enhancement of diffuse emission with respect to models of the Cygnus region 
and the inner Galaxy. 
Measuring the diffuse and extended emission in our Galaxy with better sensitivity will help us understand these enhancements, 
put tighter constraints on Galactic cosmic-ray emission, and also
distinguish between hadronic and leptonic acceleration and propagation models. 
We present in Ref.~\cite{bib:Petra} the sensitivity of the HAWC array to Galactic diffuse $\gamma$-ray emission under different model assumptions.

We also present in Ref.~\cite{bib:Crab} our observations of the Crab Nebula using data from HAWC-30, including observations during the flare in March 2013.
We expect to observe 7$\sigma$ with each transit of the Crab using the full HAWC array.  
We will measure the TeV flux of the Crab to better than 20\% for each day of the year,
enabling long-term studies of the Crab at TeV energies.

\subsection{Moon shadow}

\begin{figure}[!ht]
  \centering
  \includegraphics[width=0.45\textwidth]{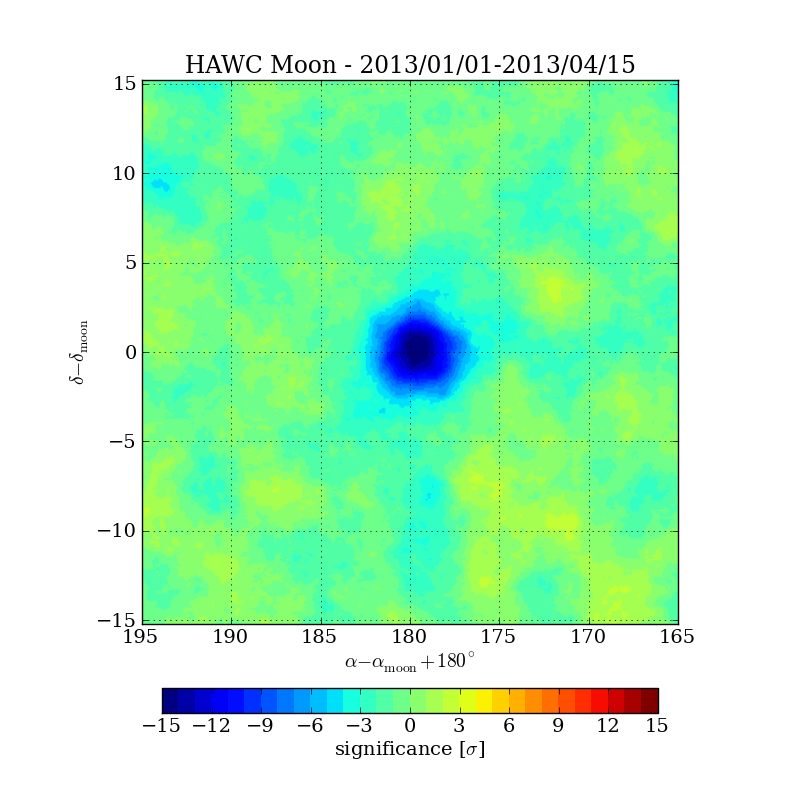}
  \caption{Skymap around the region of the Moon for 95 days of HAWC-30 data 
  in Moon-centered equatorial coordinates.
  The mapping technique, simulations, background estimation, and event selection are explained in Ref.~\cite{bib:Dan}.}
  \label{fig:moon}
 \end{figure}

The Moon causes a detectable deficit on top of a nearly isotropic flux of cosmic rays incident at Earth. 
The measurement of the angular width and position of this ``shadow" allows us to infer the accuracy of our angular reconstruction. 
We show in Fig.~\ref{fig:moon} an observation of the Moon shadow using 95 days of HAWC-30 data.
%

We report in Ref.~\cite{bib:Dan} the detection and study of the moon shadow and its deflection due to the geomagnetic field. 
Both the width and center of the observed Moon shadow are consistent with the predictions obtained from simulations.
The measurement of the amount of suppressed cosmic ray flux (in a circular region of radius 5$^\circ$ around the Moon) 
also matches the expectations. 
We infer a point-spread function of 1.2$^\circ$ for cosmic ray data from HAWC-30
by comparing the observations with simulations. 
We expect the Moon shadow position and width as well as the angular resolution of the detector to be a function of energy. 
We estimate the median energy of this subset 
to be $\sim3$~TeV. 
%

As HAWC-111 starts operation in August 2013, 
the observation of the Moon shadow will become a daily monitoring tool.
We will also be able to measure the angular resolution as a function of energy with the larger data set from a larger array.

\subsection{Transient events}

Observations of energy spectra of $\gamma$-ray bursts (GRBs) can provide information 
about the intervening space between the burst and Earth as well as about the source itself.
The HAWC observatory will be almost two orders of magnitude more sensitive to GRBs at $100$~GeV 
than the Milagro experiment due to the higher altitude. 
Due to the wide instantaneous field of view ($\sim2$~sr) and large duty cycle of the HAWC observatory, 
we will be able to observe the beginning of the prompt phase of GRBs without needing to slew. 
We will also be able to constrain the shape and cutoff of the GRB spectra in the sub-TeV to TeV energy range,
especially in conjunction with observations from other detectors such as the Fermi satellite. 

\begin{figure}[!ht]
  \centering
  \includegraphics[width=0.45\textwidth]{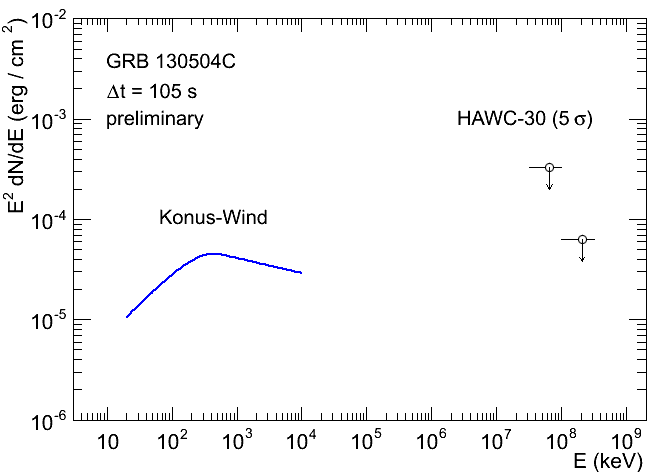}
  \caption{The $5\sigma$ upper limit on the high-energy emission from GRB~130504C imposed with HAWC-30 data. 
  The $5\sigma$ sensitivity is reported rather than $90\%$ C.L.\ upper limits to allow the data to remain blinded. 
  The spectral fit reported by Konus-Wind 
  is shown for comparison.}
  \label{fig:GRB}
 \end{figure}

GRB~130504C is the brightest GRB that occurred in the field of view of HAWC-30 during its uptime. 
A preliminary analysis using the main DAQ data for GRB 130504C shows no excess larger than $5\sigma$
in the time windows examined ($73.2$, $105.0$, and $219.6$~s). 
We show in Fig.~\ref{fig:GRB} the upper limits derived from Monte Carlo simulations for a 5$\sigma$ discovery of GRB~130504C 
using data from HAWC-30 
(for the $105$~s time window). 
This analysis will be updated to show 90\% confidence level upper limits once the data are unblinded. 
The Konus-Wind spectral fit from GCN circular 14587  is also shown in Fig.~\ref{fig:GRB} for comparison. 
More details on this analysis can be found on Ref.~\cite{bib:GRB}.

HAWC data are collected by two 
acquisition systems (DAQs). 
Using the main DAQ, we measure the arrival time and the time over threshold (TOT) of the PMT pulses.
This system provides us with the information to reconstruct the shower core, the arrival direction, and the lateral distribution.
This in turn helps us determine the species of primary particle and its energy. 
We discuss in more detail the GRB results using the main DAQ in Ref.~\cite{bib:GRB}.

The other DAQ, the scalers system, operates in a counting mode, 
and it is sensitive to $\gamma$-ray and cosmic ray 
transient events that produce a sudden increase 
or decrease in the counting rates with respect to those produced by atmospheric showers and noise. 
We present 
the analysis procedure for data 
taken with the scalers DAQ, 
and the first results for selected GRBs 
in Ref.~\cite{bib:icrcscalers}.
Although we see no indication of emission above $10$~GeV, 
GRBs like 090510 and 090902B would be significantly detected with the scalers 
if they occurred at favorable zenith angles.

The sensitivity of the HAWC observatory to GRBs increases dramatically as the array grows. 
We show in Ref.~\cite{bib:GRB} the effects of different GRB emission spectra on the expected sensitivity of HAWC 
for several stages of the deployment. 
The HAWC observatory stands an excellent chance of seeing GRBs that are detected at lower energies 
%
even before the array gets completed by the end of 2014.

The 
HAWC observatory will provide a nearly continuous, unbiased survey of TeV emission from the northern hemisphere,
which makes it ideally suited for detecting bright transient events, such as outbursts from active galactic nuclei (AGN). 
%
We are implementing a comprehensive, online flare-monitoring program to promptly detect bright outbursts from moderate-redshift AGN. 
This will provide unique opportunities to trigger follow-up observations at complimentary wavelengths. 
We discuss in Ref.~\cite{bib:Asif} the various components of the flare-monitoring system. 

\section{Observations of cosmic rays}

The main goal of the HAWC observatory is to study $\gamma$-rays
between $50$~GeV and $100$~TeV.
Although charged cosmic rays are the major source of background, 
the distribution of the arrival directions of the cosmic rays in this energy range is of significant interest to the particle astrophysics community.  
%
An anisotropy in the distribution of arrival directions of TeV cosmic rays has been 
observed  at the $10^{-3}$ level on large ($>60^\circ$) and
small ($<20^\circ$) angular scales by multiple experiments
in both hemispheres.
The large-scale structure
is dominated by dipole and quadrupole moments and does not appear to persist
above the TeV energies.  
Although many explanations have been suggested,
the origin of the observed anisotropy is not understood.

We measure the cosmic-ray anisotropy in the TeV energy range  
using data 
collected during the operation of HAWC-30 between January 1, 2013 and April 15, 2013.
(This corresponds to a live-time of $95$~days.)
We require a 
zenith angle below $45^\circ$ and at least $15$ triggered PMTs.
The rate of air showers from cosmic rays recorded with the HAWC-30 array is $>2$~kHz. 
We accumulated $2.2\times10^{10}$ well-reconstructed events during this period. 
%
We estimate a median energy of $2$~TeV for this data set.

\begin{figure*}[!hbt]
  \centering
  \includegraphics[width=0.8\textwidth]{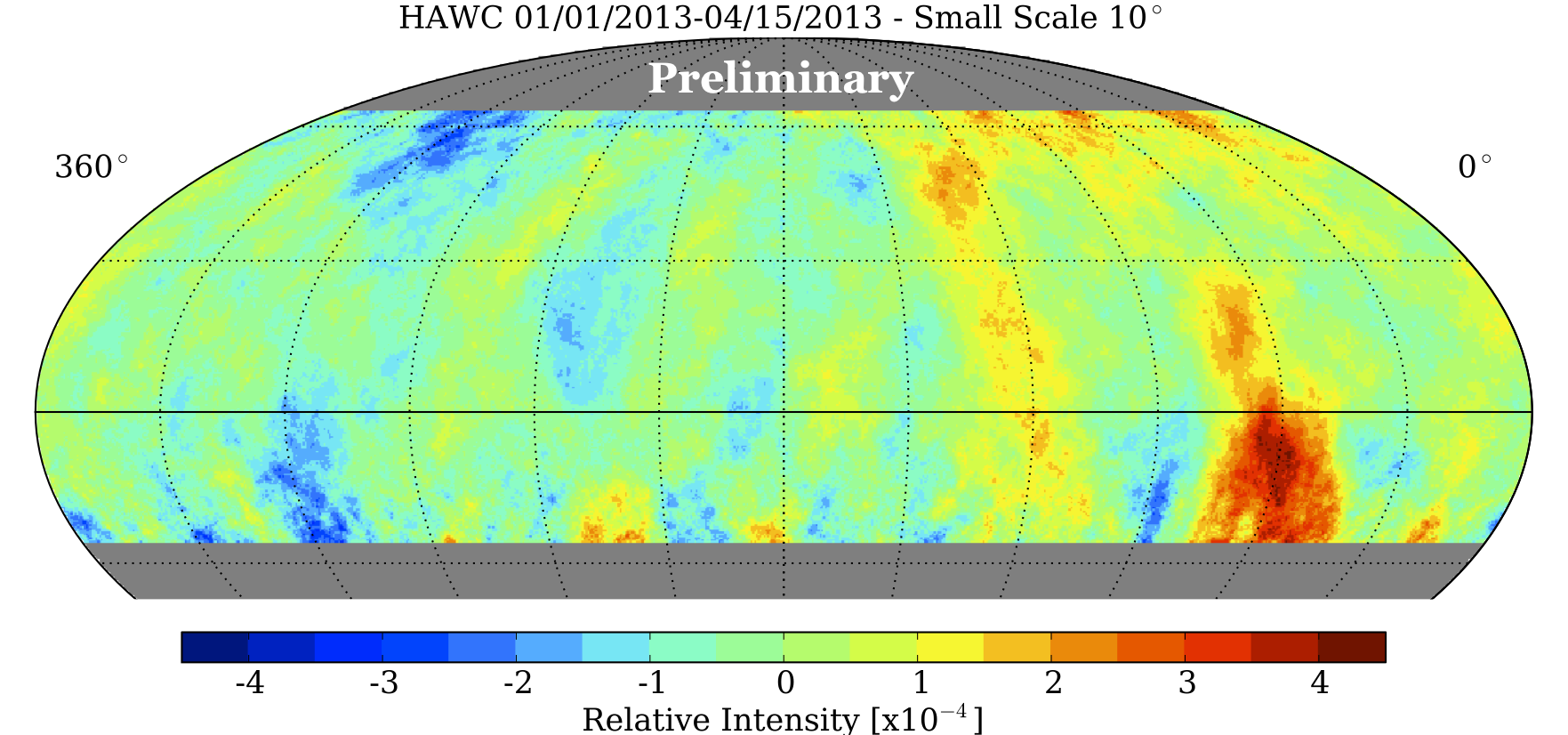}
  \caption{Relative intensity of the HAWC-30 cosmic-ray data produced using a time-integration window $\Delta t = 2$~hr.
The sky map of the pre-trial significances of the features in this map are shown in Ref.~\cite{bib:CRani}.}
  \label{fig:crani}
 \end{figure*}

To search for anisotropies on small angular scales, 
we compute 
the amplitude of the deviations from the isotropic expectation in each pixel.
We determine 
the deviations from isotropy by calculating 
the relative intensities (with respect to a reference map) as a function of equatorial coordinates.
We show the measured relative intensities in Fig.~\ref{fig:crani}.
Data have been smoothed using a $10^\circ$ top-hat function. 
Our measurements are sensitive 
to features smaller than $30^\circ$ in right ascension
because we use an integration time of $\Delta~t=2$~hr to generate the maps.
Due to the low latitude of the HAWC site, 
these data cover a region of the sky previously unobserved by other experiments. 

We present in Ref.~\cite{bib:CRani} the results of the search for anisotropy on large and small angular scales
using data from HAWC-30. 
We also compare the observed anisotropy with previous measurements of the northern and southern skies.
Several prominent features are visible in the sky map, notably the regions of
excess flux at $\alpha=60^\circ$ and $\alpha=120^\circ$.
There are of the order of $10^5$ independent pixels in the sky map, and after
accounting for trial factors only these two regions of excess are significant at a level larger than $5\sigma$.  
These hot spots correspond to the $10^\circ$--~$20^\circ$ regions of cosmic ray excess previously observed by Milagro 
and ARGO. 
%
We also note that
regions of excess are associated with neighboring deficits in the cosmic ray flux.  
We cannot currently rule out the possibility that the deficit regions are artifacts of the analysis 
rather than real features in the residual cosmic ray flux.

\subsection{Solar events}

Although one of the primary purposes of the HAWC collaboration 
is to study 
sources of high energy $\gamma$-rays,
the counting rate of the HAWC observatory will be sensitive to cosmic rays with energies above the 
geomagnetic cutoff of the site ($\sim 9$~GV). 
In particular, we will be able to detect solar energetic particles 
and the effect of coronal mass ejections on the galactic cosmic rays. 
This effect is known as Forbush Decreases (FDs). 
We present in Ref.~\cite{bib:Lara} 
the capabilities of the HAWC observatory to observe 
solar energetic events,
and the observation of an FD in the data from HAWC-30.

\begin{figure}[!ht]
  \centering
  \includegraphics[width=0.475\textwidth]{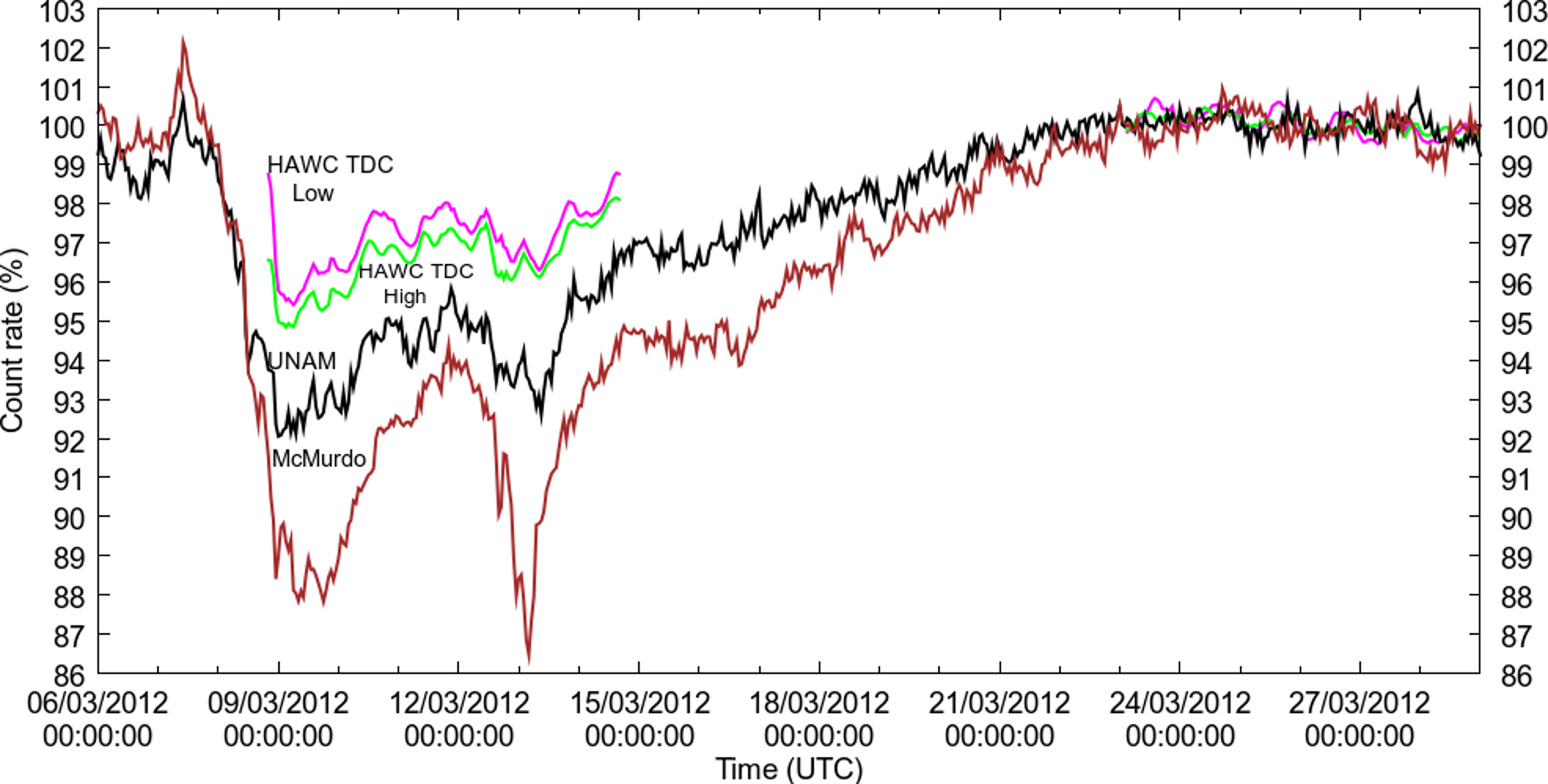}
  \caption{Comparison between the counting rate of the main DAQ system of the VAMOS prototype (in magenta for 2-edge hits and in green for 4-edge hits), 
  the Mexico City Cosmic Rays Observatory (in black), and the McMurdo station (in brown) during a Forbush decrease. 
  The  magnetic rigidity cutoffs are $9$~GV, $8.2$~GV, and $0.3$~GV, respectively.}
  \label{fig:FD}
 \end{figure}

We show in Fig.~\ref{fig:FD} the observation of an FD in the data from the engineering prototype VAMOS using the main DAQ.
We describe the analysis of data from both the main DAQ and the scalers system, and the corrections due to atmospheric effects 
in Ref.~\cite{bib:Mario}.
The observation of this transient event in the data from VAMOS is in very good agreement with 
neutron monitors located in Mexico City and the South Pole.

\section{Fundamental physics with HAWC}

The high energy coverage of the HAWC observatory
enable us to study fundamental physics beyond the Standard Model. 
The large field of view of the observatory will also allow detailed studies of cosmologically significant backgrounds and 
intergalactic magnetic fields.  
We describe in this section the sensitivity of the HAWC observatory to dark matter, cosmology, and fundamental physics.

\begin{figure}[!ht]
  \centering
  \includegraphics[width=0.475\textwidth]{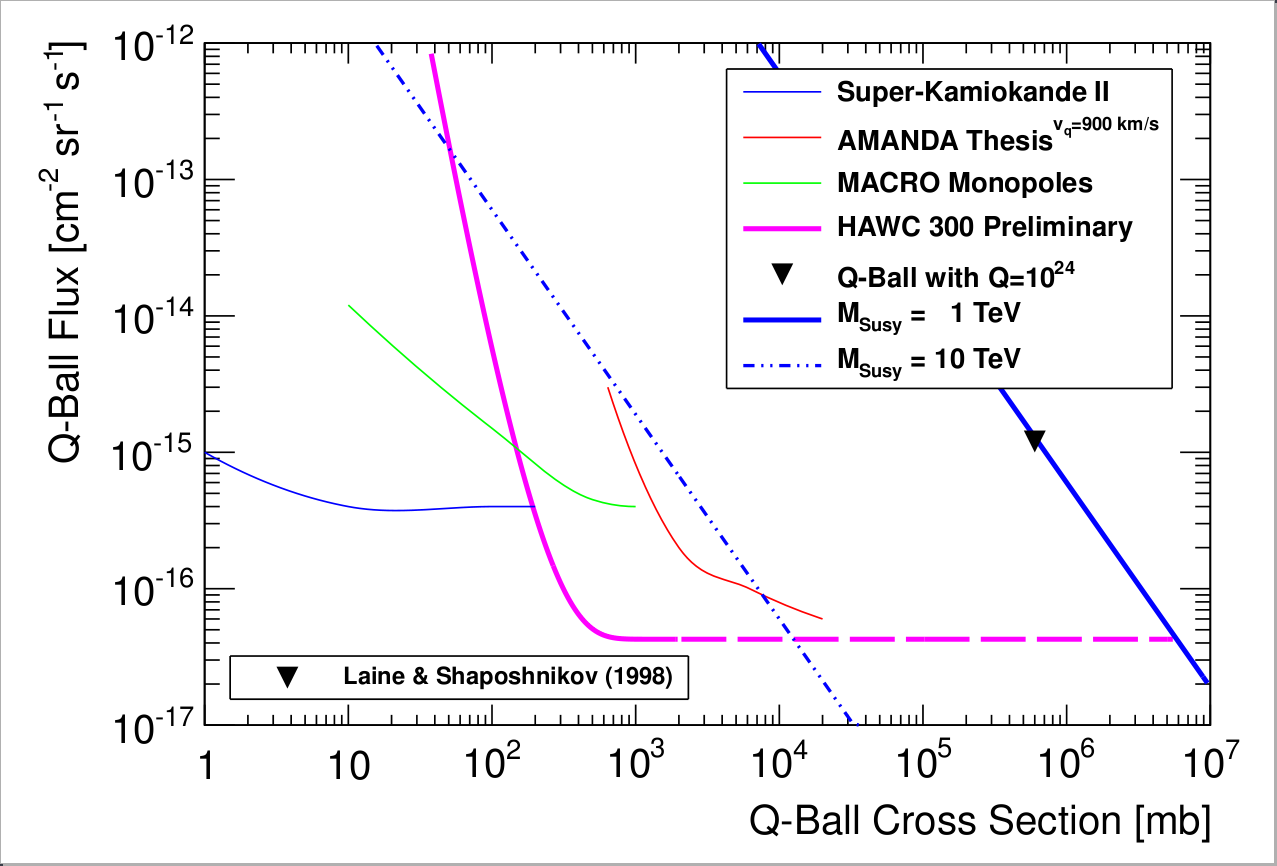}
  \caption{Estimated flux upper limit to Q-balls using one year of data of the HAWC observatory (assuming no background) compared to current limits. 
  The solid line indicates what can be done with the current Q-ball trigger algorithm while the dashed line indicates cross sections which could be accessible with a new saturation trigger.
  The blue lines on the right show expectations from theory.
  Details about this analysis can be found in Ref.~\cite{bib:Peter}.}
  \label{fig:Qballs}
 \end{figure}

Each WCD holds $\sim$200,000 liters of ultra-pure water.
Thus, the completed HAWC array will be a $60$~kton detector.
Its flat geometry also provides more effective area for a given volume.
These features of the HAWC array give us the opportunity to search for exotic particles. 
One interesting example is the Q-balls that are predicted by several supersymmetric theories.
Q-balls are very massive, sub-relativistic particles that can have a large baryon
number and can be stable since their creation in the early universe. 
They are also an appealing candidate for the dark matter of the universe, 
but their large masses imply that their flux is very low. 
The HAWC observatory has a flexible DAQ system that
allows us to search for Q-balls traversing the detector.
For this purpose, a dedicated trigger algorithm for non-relativistic species was designed and implemented.
We show in Fig.~\ref{fig:Qballs} the sensitivity to Q-balls for one year of data of 
the full HAWC array in comparison to established limits and a theoretical prediction.
For this limit we assume that background triggers can be removed after reconstruction.  
This is possible with a more restrictive event selection in the trigger algorithm.
We describe the fundamentals of Q-balls, and the dedicated trigger algorithm in Ref.~\cite{bib:Peter}.

A growing body of evidence exists supporting the existence of Dark Matter (DM), yet its particle nature remains a mystery.
Weakly-interacting massive particles (WIMPs) with a mass ranging from $10$~GeV to $10$~TeV are a popular candidate to explain 
the particle nature of DM.
A WIMP hypothesis is compelling as it naturally explains the density of DM observed, 
and such particles are expected in some solutions to the gauge hierarchy problem.
WIMPs could annihilate into Standard Model particles and produce high energy photons via various processes.
Detecting such high energy photons would provide an indirect detection of the DM particle mass.

\begin{figure}[!ht]
  \centering
  \includegraphics[width=0.5\textwidth]{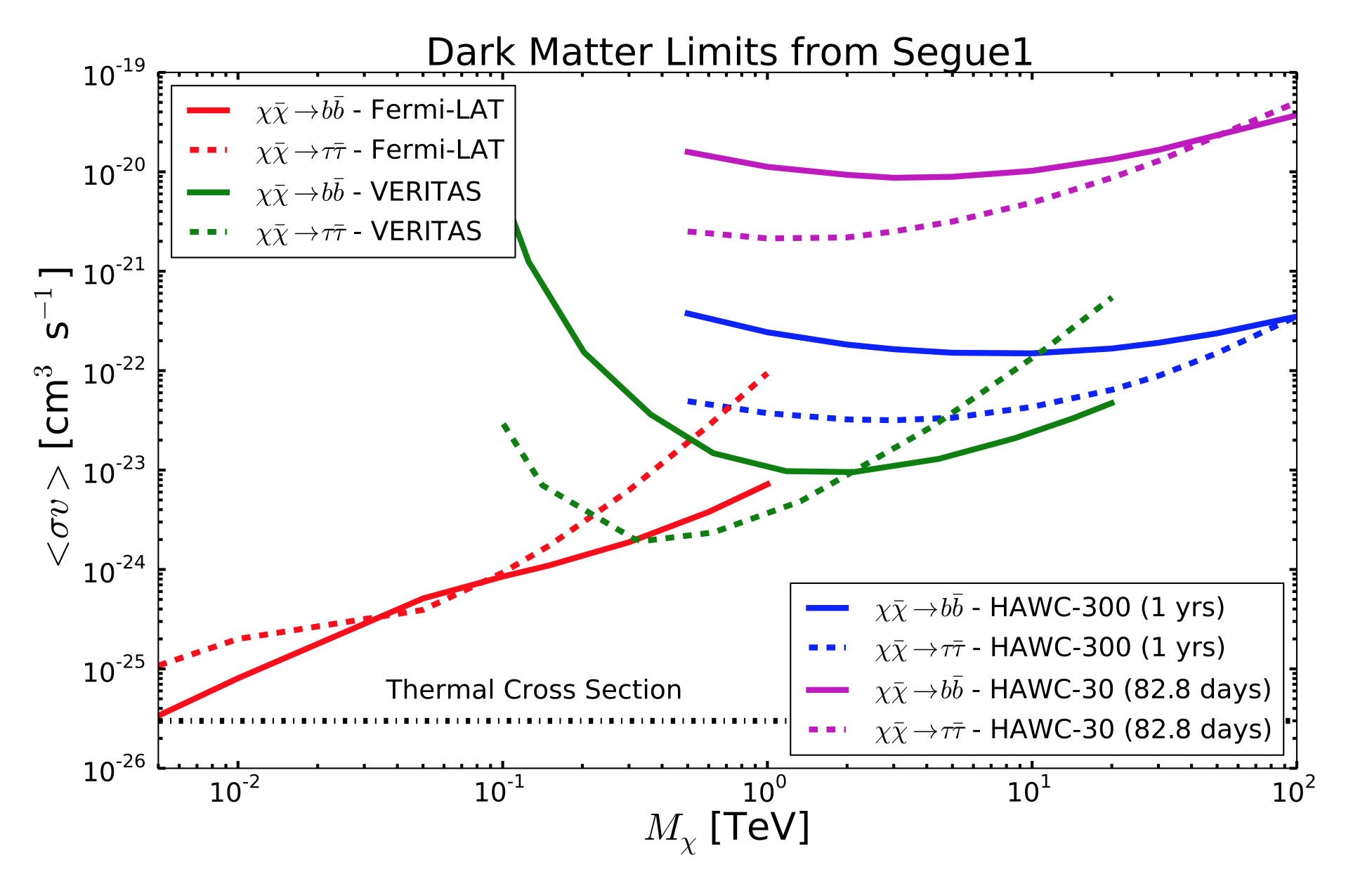}
  \caption{The preliminary sensitivity on $\left<\sigma\ v\right>$ vs.\ $M_\chi$ in the annihilation channels 
  $\chi \bar\chi \rightarrow  \tau \bar\tau$ and $\chi \bar\chi \rightarrow  b \bar b$ from observations of Segue~1 
  with HAWC-30 ($82.8$~days), and the expected sensitivity for the HAWC observatory (1~year)
  compared to the 
  current limits from Fermi-LAT (24~months) and VERITAS (50~hours).}
  \label{fig:Segue1}
 \end{figure}

The HAWC observatory can set 
stringent limits of indirect detection of WIMP candidates above $100$~GeV
due to its large effective area, excellent angular resolution, efficient gamma-hadron separation, and large duty cycle.
We show in Fig.~\ref{fig:Segue1} the preliminary limits on DM candidate velocity weighted cross-section 
as a function of the DM mass for observations of Segue~1.
We use data collected with HAWC-30 ($82.8$~days) in this analysis. 
These represent the most stringent limits available above $20$~TeV.
We also show the predicted sensitivity of the HAWC observatory after completion in 2014.

We present in Ref.~\cite{bib:Pat} 
these preliminary limits for various DM annihilation channel hypotheses.
As we continue deployment, the sensitivity of the HAWC array will increase due to improvements in effective area, angular resolution, 
and $\gamma/$hadron separation.
Observations with the HAWC observatory on DM dominated objects such as Segue~1 will provide competitive limits on 
masses above $1$~TeV. 

Primordial black holes (PBHs) are hypothetical objects that formed from extreme densities of matter 
that were present during the early universe. 
PBHs with large enough initial masses 
could be bursting high energy particles, 
including very high energy $\gamma$-rays. 
The HAWC observatory 
will be an important tool to
either detect a PBH burst or set stringent new limits on the PBH burst rate. 
We present in Ref.~\cite{bib:Tilan} the sensitivity of the HAWC observatory to PBH bursts according to the Standard Model.

Lorentz invariance is believed to be a fundamental symmetry. 
The observation of a Lorentz invariance violation (LIV) would revolutionize our view of  the universe and probe physics 
at energy scales not attainable with manmade accelerators. 
The combination of extreme distance, high energy emission, 
and short duration 
makes GRBs an ideal laboratory to search for LIV. 
We discuss in Ref.~\cite{bib:Lukas} the current limits on LIV and 
the potential to study possible deviations of the speed of photons from $c$. 
We will be able to set competitive limits on the scale of LIV 
using GRB observations with the HAWC observatory. 

The intergalactic magnetic field (IGMF), 
presumed to exist in the void regions between galaxy clusters, 
may play a role in galaxy formation, contain information about conditions in the early universe, 
and influence the trajectories of cosmic rays of extragalactic origin.
Recent studies have attempted to measure the IGMF by searching for its influence on the cascades 
produced when very high energy gamma rays from blazars interact with the extragalactic background light (EBL).
The HAWC observatory
will provide unbiased observations of the average fluxes from blazars and 
accurate measurements of the attenuation of $\gamma$-rays due to interactions with the EBL.
We present in Ref.~\cite{bib:Tom} the capabilities of the HAWC observatory to contribute to measurements of the IGMF 
from the observations of the delayed secondary flux following a bright blazar flare.

\section{Conclusions and outlook}

Deployment of the HAWC array
started in March 2012. 
After only six months we were already operating an array of 30 WCDs with a size comparable to the Milagro experiment.
The modularity of the observatory allows us to take data during deployment as the WCDs are instrumented and verified as they are placed into the data stream.
We present at this conference the first results on
the observation of the shadow of the moon, the observation of small-scale and large-scale angular
clustering of TeV cosmic rays, the measurement of transient
solar events, the observation of Forbush decreases,
upper limits on high energy emission from GRBs, 
the observation of the Crab Nebula,
a program to monitor AGN flares,
and the sensitivity of the observatory to 
Q-balls, primordial black holes, Lorentz invariance violation, and
the indirect detection of dark matter.
We will have approximately one year of data from HAWC-111 for Summer 2014.
At the next ICRC in The Netherlands, we will have the first results from the full HAWC array,
one order of magnitude more sensitive than its predecessor.

\vspace*{0.5cm}
\footnotesize{{\bf Acknowledgment:}{We acknowledge the support from the U.S.\ National Science Foundation; U.S.\ Department of Energy Office of High-Energy Physics; The Laboratory Directed Research and Development program of Los Alamos National Laboratory; Consejo Nacional de Ciencia y Tecnolog\'{\i}a, M\'exico; Red de F\'{\i}sica de Altas Energ\'{\i}as, M\'exico; DGAPA-UNAM, M\'exico; and the University of Wisconsin Alumni Research Foundation.}}


\begin{thebibliography}{}

\bibitem{bib:TeVCat} http://tevcat.uchicago.edu

\bibitem{bib:Hugo} ``Timing Calibration of the HAWC Observatory," H.\ A.\ Ayala Solares, H.\ Zhou, C.\ M.\ Hui, and P.\ H{\"u}ntemeyer 
for the HAWC Collaboration, Proceedings of the ICRC'13. 
(\href{http://arxiv.org/abs/1310.0074}{arXiv:1310.0074} [astro-ph.IM])

\bibitem{bib:Robert} ``Calibration and Reconstruction Performance of the HAWC Observatory," R.\ J.\ Lauer 
for the HAWC Collaboration, Proceedings of the ICRC'13. 
(\href{http://arxiv.org/abs/1310.0074}{arXiv:1310.0074} [astro-ph.IM])

\bibitem{bib:Ibrahim} ``Deployment of the HAWC gamma-ray observatory in Sierra Negra, Mexico," I.\ Torres, A.\ Carrami\~nana, 
R.\ Alfaro, A.\ Iriarte for the HAWC Collaboration, Proceedings of the ICRC'13. 
(\href{http://arxiv.org/abs/1310.0074}{arXiv:1310.0074} [astro-ph.IM])

\bibitem{bib:VAMOS} ``VAMOS: a pathfinder for the HAWC gamma ray observatory," The HAWC Collaboration, in preparation.

\bibitem{bib:Jordan} ``Sensitivity and Status of HAWC," J.\ Goodman, and J.\ Pretz  for the HAWC Collaboration, 
Proceedings of the ICRC'13. 
(\href{http://arxiv.org/abs/1310.0071}{arXiv:1310.0071} [astro-ph.HE])

\bibitem{bib:Michelle} ``Studying the Cygnus region with the HAWC Observatory," C.\ M.\ Hui for the HAWC Collaboration, 
Proceedings of the ICRC'13. 
(\href{http://arxiv.org/abs/1310.0071}{arXiv:1310.0071} [astro-ph.HE])

\bibitem{bib:Petra} ``HAWC Sensitivity to Diffuse Emission," P.\ H\"untemeyer, and H.\ A.\ Ayala Solares for the HAWC Collaboration, 
Proceedings of the ICRC'13. 
(\href{http://arxiv.org/abs/1310.0071}{arXiv:1310.0071} [astro-ph.HE])

\bibitem{bib:Crab} ``HAWC Observations of the Crab Nebula," B.\ M.\ Baughman, J.\ Braun, J.\ A.\ Goodman, A.\ Imran, B.\ Patricelli, and J.\ Pretz for the HAWC Collaboration, 
Proceedings of the ICRC'13. 
(\href{http://arxiv.org/abs/1310.0071}{arXiv:1310.0071} [astro-ph.HE])

\bibitem{bib:Dan} ``Observation of the Moon shadow and characterization of the point response of HAWC-30," D.\ Fiorino, S.\ BenZvi, 
and J.\ Braun for the HAWC Collaboration, Proceedings of the ICRC'13.
(\href{http://arxiv.org/abs/1310.0072}{arXiv:1310.0072} [astro-ph.HE])

\bibitem{bib:GRB} ``Search for high-energy emission from GRBs with the HAWC Observatory," K.\ Sparks 
for the HAWC Collaboration, Proceedings of the ICRC'13. 
(\href{http://arxiv.org/abs/1310.0071}{arXiv:1310.0071} [astro-ph.HE])

\bibitem{bib:icrcscalers} ``Sensitivity of the HAWC Observatory to Gamma-ray Bursts Using the Scaler System," D.\ Lennarz 
for the HAWC Collaboration, Proceedings of the ICRC'13. 
(\href{http://arxiv.org/abs/1310.0071}{arXiv:1310.0071} [astro-ph.HE])

\bibitem{bib:Asif} ``A Real-time AGN Flare Monitor for the HAWC Observatory," A.\ Imran, and R.\ Lauer 
for the HAWC Collaboration, Proceedings of the ICRC'13. 
(\href{http://arxiv.org/abs/1310.0071}{arXiv:1310.0071} [astro-ph.HE])

\bibitem{bib:CRani} ``Observation of the Anisotropy of Cosmic Rays with HAWC," S.\ BenZvi, D.\ Fiorino, and K.\ Sparks 
for the HAWC Collaboration, Proceedings of the ICRC'13. 
(\href{http://arxiv.org/abs/1310.0072}{arXiv:1310.0072} [astro-ph.HE])

\bibitem{bib:Lara} ``HAWC and Solar Energetic Transient Events," A.\ Lara for the HAWC Collaboration, Proceedings of the ICRC'13. 
(\href{http://arxiv.org/abs/1310.0072}{arXiv:1310.0072} [astro-ph.HE])

\bibitem{bib:Mario} ``Observation of the March 2012 Forbush decrease with the engineering array of the High Altitude Water 
Cherenkov Observatory,"  M.\ Castillo, H.\ Salazar, and L.\ Villase\~nor for the HAWC Collaboration, Proceedings of the ICRC'13. 
(\href{http://arxiv.org/abs/1310.0072}{arXiv:1310.0072} [astro-ph.HE])

\bibitem{bib:Peter} ``Searching for Q-balls with the High Altitude Water Cherenkov Observatory," P.\ Karn, and P.\ Younk 
for the HAWC Collaboration, Proceedings of the ICRC'13. 
(\href{http://arxiv.org/abs/1310.0073}{arXiv:1310.0073} [astro-ph.HE])

\bibitem{bib:Pat} ``Limits on Indirect Detection of WIMPs with the HAWC Observatory," B.\ M.\ Baughman, J.\ P.\ Harding 
for the HAWC Collaboration, Proceedings of the ICRC'13. 
(\href{http://arxiv.org/abs/1310.0073}{arXiv:1310.0073} [astro-ph.HE])

\bibitem{bib:Tilan} ``HAWC Sensitivity for the Rate-Density of Evaporating Primordial Black Holes," T.\ N.\ Ukwatta, J.\ H.\ MacGibbon,
D.\ Stump, G.\ Sinnis, J.\ T.\ Linnemann, K.\ Tollefson, A.\ U.\ Abeysekara, and D.\ Lennarz for the HAWC Collaboration, 
Proceedings of the ICRC'13. 
(\href{http://arxiv.org/abs/1310.0073}{arXiv:1310.0073} [astro-ph.HE])

\bibitem{bib:Lukas} ``Sensitivity of the HAWC Detector to Violations of Lorentz Invariance," L.\ Nellen for the HAWC Collaboration, 
Proceedings of the ICRC'13. 
(\href{http://arxiv.org/abs/1310.0073}{arXiv:1310.0073} [astro-ph.HE])

\bibitem{bib:Tom} ``HAWC Contributions to IGMF Studies," T.\ Weisgarber for the HAWC Collaboration, 
Proceedings of the ICRC'13. 
(\href{http://arxiv.org/abs/1310.0073}{arXiv:1310.0073} [astro-ph.HE])

\end{thebibliography}
\end{document}